# What do large-scale patterns teach us about extreme precipitation over the Mediterranean at medium- and extended-range forecasts?

Nikolaos Mastrantonas (1, 2), Linus Magnusson (1), Florian Pappenberger (1), and Jörg Matschullat (2)

(1) European Centre for Medium-Range Weather Forecast (ECMWF), United Kingdom; (2) Technische Universität Bergakademie Freiberg (TUBAF), Germany

**Corresponding author**: Nikolaos.Mastrantonas@ecmwf.int

## Abstract

Extreme Precipitation Events (EPEs) can have devastating consequences such as floods and landslides, posing a great threat to society and economy. Predicting such events long in advance can support the mitigation of negative impacts. Here, we focus on EPEs over the Mediterranean, a region that is frequently affected by such hazards. Previous work identified strong connections between localised EPEs and large-scale atmospheric flow patterns, affecting weather over the entire Mediterranean. We analyse the predictive skill of these patterns in the ECMWF extended range forecasts and assess if and where these patterns can be used for indirect predictions of EPEs, using the Brier Skill Score.

The results show that the ECMWF model provides skilful predictions of the Mediterranean patterns up to 2 weeks in advance. Moreover, using the forecasted patterns for indirect predictability of EPEs outperforms the reference score up to ~10 days lead time for many locations. Especially for high orography locations or coastal areas, like parts of western Turkey, western Balkans, Iberian Peninsula and Morocco this limit extends to 11–14 days lead time.

This study demonstrates that connections between localised EPEs and large-scale patterns over the Mediterranean extends the forecasting horizon of the model by over 3 days in many locations, in comparison to forecasting based on the predicted precipitation. Thus, it is beneficial to use the predicted patterns rather than the predicted precipitation at longer lead times for EPEs forecasting. The model's performance is also assessed from a user perspective, showing that the EPEs forecasting based on the patterns increases the economic benefits at medium/extended range lead times. Such information could support higher confidence in the decision-making of various users, e.g., the agricultural sector and (re)insurance companies.

## 1. Introduction

Extreme Precipitation Events (EPEs) pose a great threat to society, economy, and the environment, with consequences like landslides and floods (Jonkman, 2005). Especially for locations in the Mediterranean, this threat is one of the most crucial and frequent natural hazards, resulting in high economic losses, injuries, and casualties (Llasat et al., 2013, 2010). Being able to better predict when and where such events are expected to occur can support the mitigation of their negative impacts. This is becoming more crucial in the light of non-stationary climate (e.g. Hannaford et al., 2021) and ongoing climate change, resulting in the intensification of frequency and magnitude of such extremes in many locations (e.g. Alexander et al., 2006; Cardell et al., 2020; Kostopoulou and Jones, 2005; Papalexiou and Montanari, 2019; Toreti et al., 2013). An outcome that can be attributed to the increased temperatures of the planet, which lead, in agreement with the Clausius-Clapeyron relationship, to increased water vapour (Douville and John, 2020) and therefore to increased precipitation.

Recently, there is an ever increasing scientific and operational interest on sub-seasonal predictability of atmospheric variability and weather extremes, meaning from 10–15 days up to 2–3 months in advance. Such interest is driven not only by the needs of various sectors and users (e.g. agriculture, (re)insurance companies, emergency response units; White et al., 2017), but also by advancing Numerical Weather Prediction (NWP) models and related research (Magnusson and Källén, 2013), providing skilful information about atmospheric variability at longer lead times. An example is the skill of models in predicting atmospheric variability in the middle troposphere over the northern hemisphere, which has increased by about 1 day per decade (Bauer et al., 2015). The ECMWF model, for example, is now predicting weather variability 10 days in advance as well as it was predicting this variability 7 days in advance 30 years ago (relevant plot available at https://www.ecmwf.int/en/forecasts/charts/catalogue/plwww_m_hr_ccaf_adrian_ts).

Nevertheless, challenges remain at the sub-seasonal predictability of weather conditions, especially for surface extremes at fine spatiotemporal resolutions. Thus, there is ongoing interest in assessing and quantifying if and how other variables of higher forecasting reliability, can be used for indirectly informing about extremes at medium- and extended-range forecasts (Magnusson, 2019). Such a variable of reliable forecasting skill at the sub-seasonal scale, is the atmospheric flow variability at the lower and middle troposphere, usually depicted by large-scale patterns over extended domains (Lavaysse et al., 2018; Vitart, 2014). Many recent studies have analysed these interactions between surface conditions and such atmospheric variables (e.g. cold spells and geopotential height at 500 hPa: Ferranti et al., 2018; wind power and geopotential height at 500 hPa: Grams et al., 2017; wind extremes and geopotential height at 500 hPa: Thomas et al., 2021; precipitation and weather patterns generated from multiple variables: Hoy et al., 2014).

One of the most challenging variables is precipitation; not only regarding its accurate forecasting (Sukovich et al., 2014), but also in terms of correct spatial representation of the observational data (e.g. Khairul et al., 2018; Mastrantonas et al., 2019). This variable is of immense importance for many applications and users. In this regard, previous studies demonstrate the strong association of large-scale atmospheric flow patterns to precipitation and localized EPEs. Such diagnostic studies are, for example, conducted for locations in the Mediterranean, India and the UK (Grazzini et al., 2020; Merino et al., 2016; Neal et al., 2020; Richardson et al., 2018; Toreti et al., 2016, 2010; Xoplaki et al., 2012, 2004). These patterns can be used as prognostic tool for extremes and provide skilful predictions of surface weather variability at longer lead times. In a recent study, it was shown that UK-specific large-scale patterns can support the predictability of extreme rainfall at medium-range forecasts (Richardson et al., 2020b). The same patterns have also been used to infer information about the likelihood of coastal flooding (Neal et al., 2018) and of droughts (Richardson et al., 2020a) at medium- and extended-range lead times. Similar analysis is also implemented for other locations and domains, e.g., for Europe. For the latter, weather analogues of atmospheric variability over Europe have been used to infer information about precipitation (and temperature) at sub-seasonal scales (Krouma et al., 2021; Yiou and Déandréis, 2019).

We build upon previous research that quantified the EPE connections over the Mediterranean to large-scale atmospheric flow patterns, specifically designed over the domain (Mastrantonas et al., 2021). These patterns, presented in Figure 1, are significantly associated with localised EPEs, with each of them being the preferential EPE pattern at different regional subdomains. The patterns represent distinct atmospheric variability over the domain. They indicate negative anomalies over the western Mediterranean (Atlantic/Biscay/Iberian Low), and the eastern Mediterranean (Balkan/Black Sea Low), positive anomalies over the whole domain (Mediterranean High), or, finally, non-anomalous conditions over the entire domain (Minor High, Minor Low). That study showed that the preferential pattern at each location has a conditional EPE probability that is over 3 times higher than the nominal EPE probability for most locations; a connection that is even stronger for coastal areas and high orography locations.

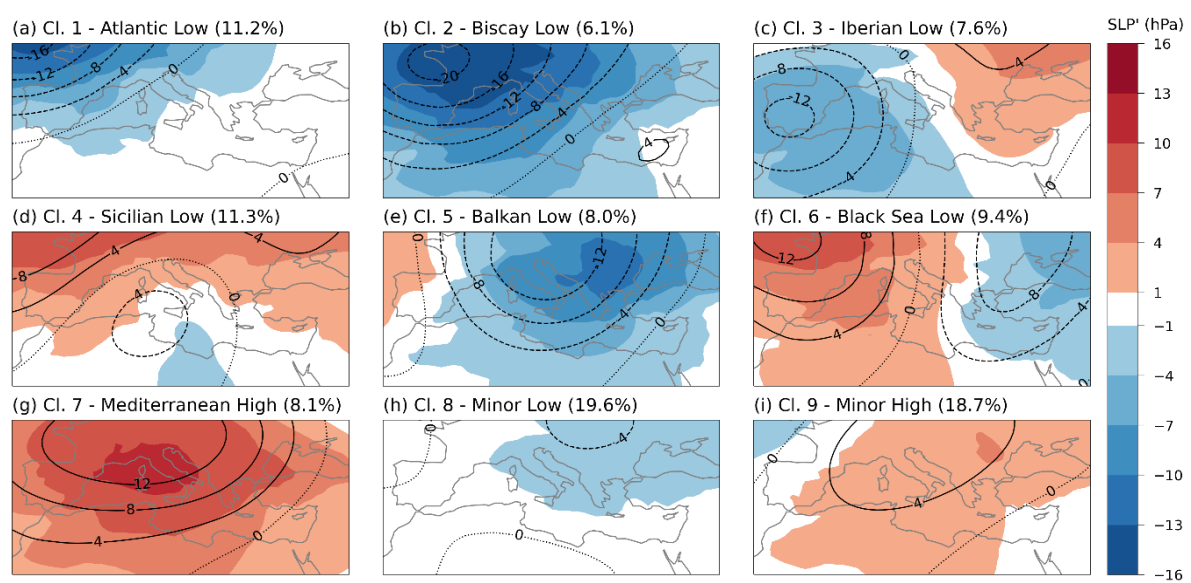

Figure 1 The 9 Mediterranean patterns (Mastrantonas et al., 2021). The figure presents the composites of clusters derived with K-means clustering on the principal components' projections of SLP and Z500 anomalies. Colour shading refers to SLP anomalies (hPa), and contours to Z500 anomalies (dam). Percentages indicate the climatological frequencies of each cluster. This figure incorporates the modifications introduced in the current study (methodology section). The differences from the original patterns and their frequencies (Fig. 5 in Mastrantonas et al., 2021), are negligible.

Based on these previous findings, this study aims at analysing the predictability of these patterns and at quantifying, if and where the predicted patterns can be used to indirectly infer useful information about EPEs. The key research questions that this study addresses are:

- What is the skill of ECMWF state-of-the-art NWP model at predicting large-scale variability over the Mediterranean at medium- and extended-range forecast scales?
- Are there systematic biases for different lead-times, seasons and patterns?
- Can forecasting of these patterns provide useful information about the indirect predictability of EPEs over the Mediterranean at medium- and extended-range forecasts?

The remaining part of this study is organized as follows: Data and methods used are described in sections 2 and 3, respectively. Section 4 presents the results and discusses the findings, and section 5 summarizes the main conclusions and suggests possible pathways for future research.

## 2. Data

As reference dataset, we used ERA5 (Hersbach et al., 2020), the latest reanalysis dataset of ECMWF for the years 1979–2020, with complete data record. ERA5 provides data at hourly resolution in a horizontal grid of roughly 30 km x 30 km, using 137 vertical levels from the surface up to 80 km height to resolve the atmosphere. The atmospheric variables used in this study are: Total precipitation, sea level pressure (SLP), and geopotential height at 500 hPa (Z500). We aggregated the data to daily resolution by summing the hourly data of the extensive variables (precipitation) and averaging the hourly data of the intensive variables (SLP, Z500).

Total precipitation was analysed at a horizontal resolution of 0.25° x 0.25° to derive statistical information of EPEs at fine spatial scales, with the selected domain covering the area 27°/48°N and 10°/41°E, referred to as **Mediterranean** from now on. Although reanalysis data have challenges in accurately representing EPE magnitudes, previous analysis has shown that ERA5 is reliable in representing EPE timing over the Mediterranean (Mastrantonas et al., 2021). Thus, given its high spatial coverage and consistency throughout the domain, it can be considered an informative source for this study. For SLP and Z500, the selected horizontal resolution was 1° x 1°; the domain covered the area 26°/50°N and -11°/41°E, so that the analysis captures the influence of adjacent areas to the Mediterranean, e.g., Atlantic Ocean and Alps.

The forecasting product used is the ECMWF extended range product (Owens and Hewson, 2018), which is run twice/week (Mondays and Thursdays) and provides information from 0 up to 46 days lead time. We used all available data produced with cycle 46r1, with the initiation dates extending between 11 June 2019 and 30 June 2020 (110 dates in total), to provide consistency in model physics and parametrization schemes. As we were interested in long-term statistical analysis, we made use of the reforecasts of these dates. Each reforecast provides ensemble data of 1 control and 10 perturbed members, for the same day-month as the actual forecasts but from 1 up to 20 years in the past. Thus, the total dataset used, consisted of 2,200 initiation dates per lead-time (20 years of reforecasts X 110 dates), with each date having 11 ensemble members. The terms "reforecasts" and "forecasts" are used interchangeably for the remaining sections and refer to the ECMWF reforecasts. The selected horizontal resolutions and domains of the three variables are the same as with ERA5. The daily values used for the analysis were derived by considering two timesteps, the 00:00 UTC of the date of interest, and 00:00 UTC of the following date.

### 3. Methodology
#### a. Large-scale patterns

The large-scale patterns in this study were derived by Mastrantonas et al. (2021), using the ERA5 data between 1979–2019. The analysis was based on Empirical Orthogonal Function (EOF) and subsequent K-means clustering of the daily SLP and Z500 anomalies. The necessary number of modes (principal components) from EOF analysis that explain at least 90% of the total variance was kept, meaning 7 and 6 for SLP and Z500, respectively. Each day was allocated to one of nine clusters, whose composites represent the nine Mediterranean patterns. Domain and variables used for generating the patterns, as well as the number of clusters, were selected so that the derived patterns have a strong association with localized EPEs over the Mediterranean and exhibit distinct synoptic-scale features over the domain. More information about the exact methodology is available at the relevant paper.

Here, we introduced a small refinement compared to Mastrantonas et al. (2021), so that the patterns can be used for operational purposes. Each day is allocated to the pattern with the minimum aggregated Euclidian distance from the 9 composites, considering the distances of both SLP and Z500. For each day, there are 9 Euclidian distances from the composites based on SLP, and 9 distances based on Z500. The timeseries of these distances for all ERA5 daily data between 1979–2019 were normalized for each variable by dividing with the mean distance of each variable from the whole dataset. Then, the aggregated Euclidian distance was calculated for each day by averaging the normalized Euclidian distances from SLP and Z500 composites. It should be noted that similar methods for allocating daily forecasts to predefined patterns based on minimum Euclidian distance is a practice implemented for other weather regimes, too, especially operationally, e.g., the 4 EuroAtlantic regimes at ECMWF (Ferranti et al., 2015), and the 30 regimes used by the Met Office (MO30; Neal et al., 2016). This method gave an initial overlap of 97% with the allocations based on the K-means clustering, indicating that both methods provide almost identical results.

b.  Connections of large-scale patterns to extreme precipitation

The connection between patterns and EPEs was quantified with the conditional probability of observing EPEs at each grid cell given each of the nine patterns. We analysed the conditional probabilities of 90$^{th}$, 95$^{th}$, and 99$^{th}$ percentile extremes (P90, P95, P99 respectively) considering full-year statistics for the period 1979–2020. Due to the high seasonal variation of the pattern frequencies, we also apply half-year statistics, with summer-halves referring to 16 April – 15 October (including both dates), and winter-halves the remaining dates, respecting the patterns' climatology (Fig. 2). Note that EPEs are always derived based on full-year statistics; the only difference is that the conditional probabilities are refined to winter- and summer-halves in addition to the pattern conditioning. Finer subsetting (e.g., seasonal) was not implemented, as the sample size for some patterns became too small, and conditional probabilities had high fluctuations (a conclusion derived after implementing bootstrapping).

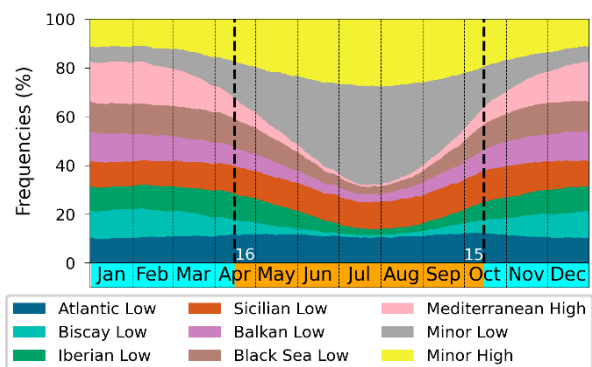

*Figure 2 91-day running mean calendar day climatological relative frequency of the 9 Mediterranean patterns. The dates of the summer-half (16 April – 15 October, inclusive of both dates) and winter-half (remaining days) periods were selected, so that the patterns' climatology is respected.*

To assess the discriminatory skill of inferring information about EPEs given the allocated pattern, the Brier Score (BS; Bouallègue et al., 2019; Wilks, 2011) was used. Each day was assigned an EPE probability at each grid cell equal to the conditional probability of the relevant pattern. The analysis was implemented for both full-year and winter-/summer-half statistics. This score was compared (Brier Skill Score; BSS) with a reference score of inferring information about EPEs based on the climatological EPE occurrence. More specifically, we used a seasonal climatology and a 31-days moving-window (centered at the date of interested) climatological occurrence of EPEs. We considered the minimum of both as the reference score. This discriminatory skill informs about if, in which locations, and how much, the use of large-scale patterns for informing about EPEs can outperform the reference score if we had a perfect forecast of the Mediterranean patterns.

c.  Skill of the ECMWF model

Initially, the model anomalies were derived after subtracting the lead-time dependent model climatology. These anomalies were used for allocating each day to one of the nine Mediterranean patterns, based on the minimum aggregated Euclidian distance, as in ERA5. The SLP and Z500 Euclidian distances of the reforecasts were normalized by dividing with the previously calculated ERA5 1979–2019 mean distance of each variable. As each reforecast day is allocated to one of the nine patterns, the BS was used to assess the performance of the model in predicting the patterns. The results were compared with the minimum of two reference scores; the BS based on 91-days moving window climatology (centered at the date of interest), and the BS based on half-year transition probabilities

(Markov chain). Moreover, the long-term frequencies of the patterns for each lead time were compared with the frequencies of these dates based on ERA5 data for assessing possible biases in pattern allocations of the ECMWF model. To assess the results' significance, we used bootstrapping of 1,000 resamples with replacement, each of them having 2,200 dates (as the size of actual sample). We considered 90% two-tailed confidence interval for the frequencies, and 90% one-tailed confidence interval for BSS. We also performed the analysis independently for winter- and summer-halves. Block-bootstrapping was not used, although it is of importance due to the high temporal correlation of the atmospheric variables. This is because the ECMWF extended-range reforecasts are not produced daily, but twice per week.

The skill of the model in predicting EPEs (P90, P95, P99) was assessed also based on BS. We calculated the direct BS for EPEs when considering the model-predicted precipitation, and the indirect BS when addressing the model-predicted patterns and their connections to EPEs. The latter was derived by substituting each predicted pattern with the relevant conditional probabilities for EPEs at each grid cell. As each reforecast has 11 members, the final conditional probabilities used for the indirect BS were the average of all 11 members at each grid cell. Thus, the maximum forecast probability that can be provided with the indirect method is the maximum conditional probability of EPEs, which occurs when all ensemble members indicate the most preferential pattern for EPEs at a specific grid cell. Note that the conditional probabilities of EPEs given the patterns (and the climatological occurrences) were derived based on the overall 1979–2020 ERA5 data. To assess the significance of the results (BSS>0), we used bootstrapping of 1,000 resamples with replacement and considered 90% one-tailed confidence interval. To obtain a larger sample of data at each bootstrap (the actual set had 2,200 at each lead time), and to respect the full-year statistics that EPEs are based on, each resample had 3,000 dates, with climatologically stable number of winter, spring, summer, and autumn days (741, 756, 756, 747 days respectively). For each resample, we finally calculated the BS for EPEs, given a perfect forecast of the patterns, so we could assess the statistical significance of the discriminatory skill presented in the previous section of the methodology.

### d. Relative economic value of forecasts

One way to look at the forecasts' skill from a user perspective, is to calculate the relative Economic Value (EV) of the forecasts. This indicator informs users about the economic benefits that could be obtained by basing the decision on the outputs of a forecasting system rather than climatological information. The EV is calculated for all possible cost-loss ratios of a preventing action, ranging from 0 to 1, so that any user can assess the usefulness of the system for their operational needs and constraints. It is a function of the frequency of the studied event, its cost-loss ratio and the false-alarm rate and hit rate, with the latter two referring to the skill of the forecasting model. For an ensemble model, this value is in fact the maximum possible, when making optimal use of the model and selecting the most appropriate probability threshold for converting the probabilistic forecast to dichotomous. This threshold depends on the cost-loss ratio. Richardson (2000) provides a comprehensive analysis about the complete sets of formulas and explanations for deriving this indicator. Here, we present the main formula:

$$EV = \frac{EV_{climatology} - EV_{forecast}}{EV_{climatology} - EV_{perfect}}$$

EV follows the logic of a skill score with positive values indicating that the forecast brings larger economic benefits compared to climatology. The upper limit is 1 and is achieved with a perfect knowledge of the future. This formula was used for understanding the economic benefits of EPE forecasting based on the ECMWF model; either direct forecasts, when considering the forecasted precipitation, or indirect forecasts, when considering the forecasted patterns and their connections to EPEs.

## 4. Results

Here, the results of P95 EPEs are presented, as they correspond to a good trade-off between extremity and sample size. See the supplementary data for results on P90 and P99 EPEs.

Figure 3 presents the pattern that corresponds to the highest conditional probability of P95 EPEs for each grid cell of the studied area. Results refer to the full-year statistics (1$^{st}$ column), and winter- and summer-half subsetting (2$^{nd}$ and 3$^{rd}$ column respectively). As mentioned in the methodology, EPEs are always derived based on full-year statistics; the only difference is that the conditional probabilities are refined to winter- and summer-halves in addition to the pattern conditioning for the two latter columns. This temporal subsetting is in fact supported by previous studies demonstrating that there are different precipitation processes active in summer versus winter (e.g. Grazzini et al., 2021). In general, each pattern is preferentially associated with EPEs at different subdomains. For example, the Biscay Low is the main EPE pattern in parts of Morocco, Iberian Peninsula, France, Italy and western Balkans, while the Black Sea Low mainly affects locations in Turkey. These results can be explained by the air- and subsequent moisture-flow associated with each of the patterns' composites (Fig. 1). Results between the 3 temporal subsets do not vary much; the main differences are identified in the Middle East. These differences are not that crucial for EPEs, as the associated probabilities are very low for these locations and temporal subsets (Fig. 3, 2$^{nd}$ row), meaning that EPEs are generally not expected in these periods and grid cells (in fact most EPEs over the domain occur in winter-halves; Mastrantonas et al., 2021). In general, the conditional probabilities of EPEs, given the most preferential pattern at each grid cell, are about 3 times higher compared to the climatological ones (5 % for P95 EPEs). Especially for locations of high orography and coastal areas, this ratio is even higher (~ 5 times). The half-year subsetting drives an increase of the conditional probabilities, as it takes the temporal occurrence of EPEs into consideration. For most of the domain, the majority of EPEs occurs in winter half years. Thus, the conditional probabilities are higher for this period. Summer half is the preferential period for EPEs in north Balkans and Alps; thus, the conditional probabilities are higher for that period for these two regions. Similar conclusions can be extracted for P90 and P99 EPEs (Figs. S1, S2).

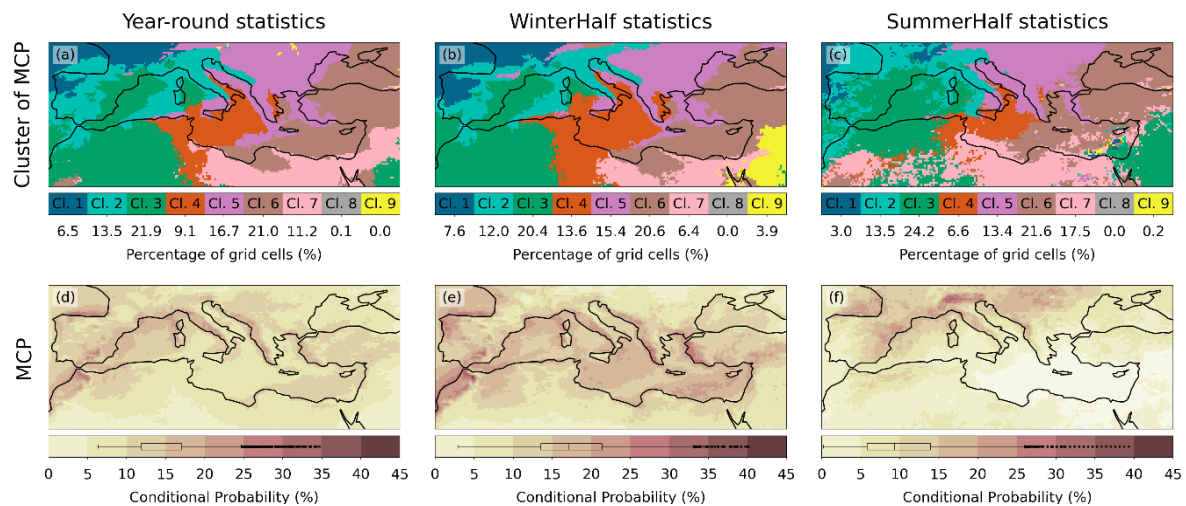

*Figure 3 Connection of P95 EPEs and Mediterranean patterns. The 1$^{st}$ row presents the pattern of Maximum Conditional Probability (MCP) at each grid cell; the 2$^{nd}$ row presents associated probabilities. The 1$^{st}$ column is based on full-year conditional probabilities; the 2$^{nd}$ and 3$^{rd}$ are based on the half-year periods (winter- and summer-half years). Note that the EPEs are always derived based on annual analysis. The boxplots at subplots (d)–(f) present the distribution of the conditional probabilities for all grid cells, indicating the median value and extending from the lower to upper quartile. The whiskers extend to the further available value up to 1.5*IQR from the lower and upper quartile; all other values outside this range are presented as outliers.*

Figure 3 presented the conditional probabilities for the preferential cluster. When using these 9 patterns for inferring information about EPEs, the discriminatory skill that can be derived for long-term EPEs analysis is most interesting (Fig. 4 for the P95 EPEs). The left plot presents the BSS when using full-year conditional probabilities. Results are masked and only present locations where the discriminatory skill outperforms the reference score. In general, conditioning based on the patterns beats the reference score in most locations except for the south-eastern Mediterranean, where EPEs are very seasonal. When further refining the conditional probabilities, based on half-year temporal subsetting (right subplot), the use of patterns outperforms reference score almost everywhere; for most locations there is more than 2.5% increase in BSS. This is the best possible improvement given a perfect knowledge (forecast) of the Mediterranean patterns. As with Figure 3, locations of high orography and coastal areas have increased performance (over 7% skill increase). This can be explained by the strong influence of topography in the EPE generation. A topography that can be considered a "passive mechanism of forced convection". The latter makes it more common to observe EPEs under specific patterns and large-scale flows that direct air and moisture perpendicular to orography. This is not true for locations over the sea and northern Africa, where topography is generally smooth over large areas (except for Atlas Mountains). Northern Africa, moreover, lies under the sinking air masses of the Hadley cell, which largely minimizes the formation of large-scale unstable conditions. Thus, EPEs in these regions mostly show more localized drivers compared to the northern Mediterranean. For the Middle East, the conditioning does not bring many benefits, as none of the 9 patterns is strongly related with EPEs over that domain. One reason is the Cyprus Low circulation that drives many of the EPEs in that region (Toreti et al., 2010), and which is not identified by the 9 patterns, due to its lower spatial extent on the studied domain (Mastrantonas et al., 2021). Similar conclusions can be extracted for P90 and P99 EPEs (Figs. S3, S4). These figures show that the discriminatory skill decreases with lower thresholds, which can be explained as follows: convective EPEs, which can be of high magnitude and are thus very rare, are mainly driven by localized characteristics, meaning that the association with large-scale patterns is not very strong.

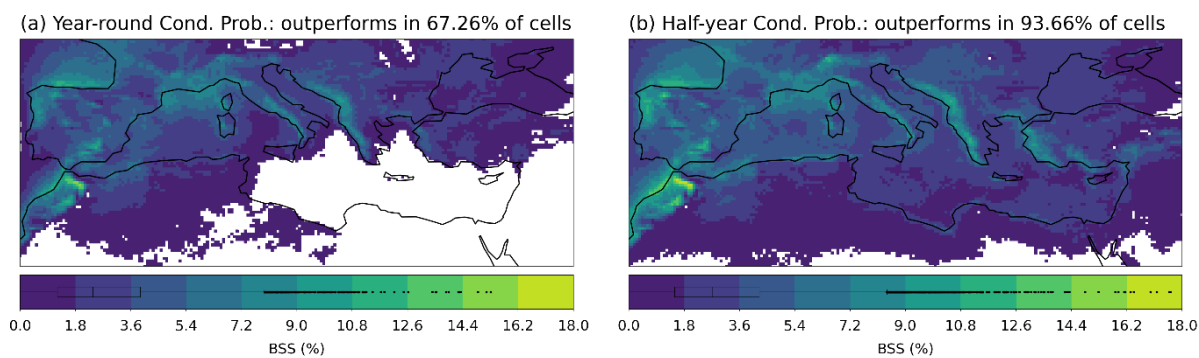

*Figure 4 Discriminatory skill (Brier Skill Score; BSS) to infer P95 EPEs, given the 9 Mediterranean patterns. Subplot (a) presents the results when using annual conditional probabilities, and subplot (b) shows the results based on the patterns and considering the half-year conditional probabilities (meaning 18 conditional probabilities in total: 9 patterns x 2 seasons; winter-/summer-half). The subplots are masked and only present the areas where the discriminatory skill outperforms the reference score (minimum of 31-day moving window climatological occurrence, and seasonal occurrence of EPEs). Boxplots as in Figure 3, but only for non-masked grid cells.*

So far, we presented the analysis of connecting EPEs and large-scale patterns based on reanalysis data. The main interest of this work is the usefulness of these connections for medium- and extended-range forecasts. Initially, Figure 5 presents the performance of the ECMWF model (BSS) in predicting the nine patterns (median value of the bootstraps). The points, connected with bold lines, indicate that the model significantly outperforms the reference scores (90% 1-tailed confidence interval). The model performance is similar for all patterns and both temporal subsets, with a forecasting horizon of about 11 days. The main difference is observed for the winter half and the patterns Mediterranean High and Minor Low. The former has a forecasting horizon of over 2 weeks, while the latter is constrained up to 5 days only. Biscay Low, a pattern highly associated with EPEs, is also more predictable during winter half. These results agree with other studies that analyse pattern predictability over other domains. Richardson et al. (2020b) showed that the ECMWF 51-member model has a predictability limit of about 15 days for the 8 UK-defined regimes, when analysing the performance based on Brier Score and considering a 1-day flexibility window in the occurrence of the patterns. Increased model skill is also observed in this study, when considering flexibility window in the occurrence of the patterns (not shown). To obtain a more comprehensive understanding about model performance, Figure S5 presents the decomposition of the Brier Score. The resolution saturates at about 11 days lead time, while the reliability has high fluctuations, especially so in winter half and annual statistics. These results also relate to the climatological pattern frequencies (Fig. 2), with high impact on the uncertainty component of the BS decomposition.

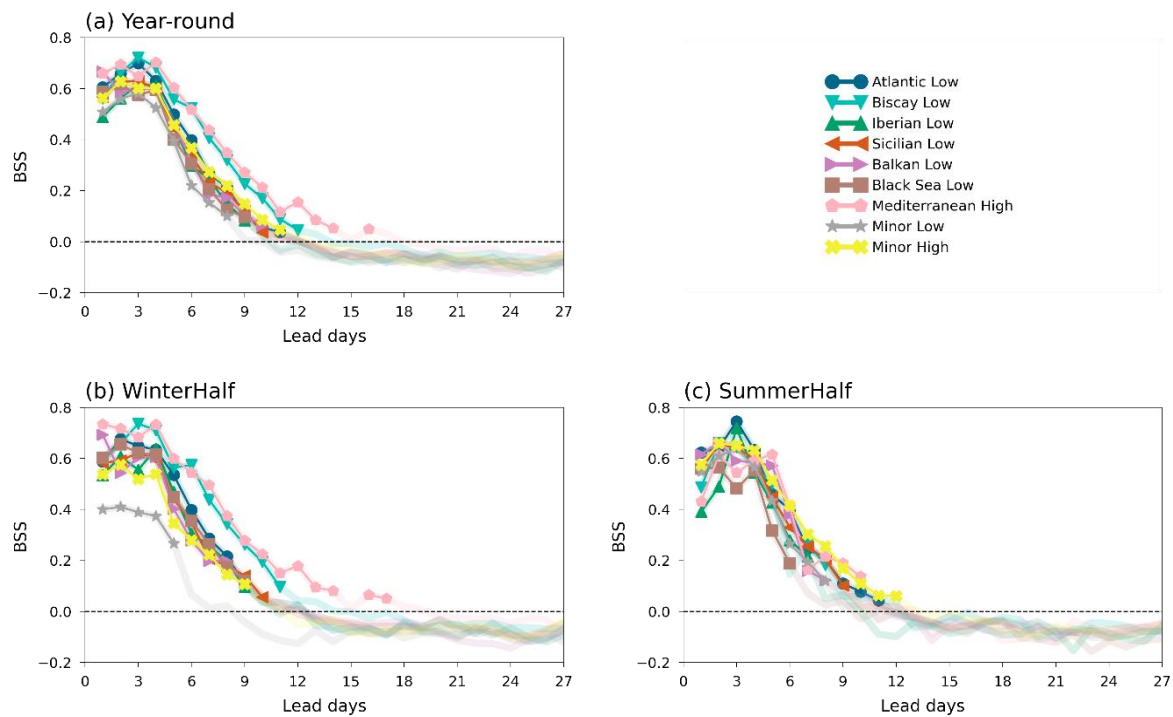

*Figure 5 Brier Skill Score (BSS) for the 9 Mediterranean patterns (median value of the bootstraps), considering the annual (a), winter-half (b), and summer-half (c) periods. The points (connected with bold lines) indicate the lead times that the model significantly outperforms the reference score.*

Figure 6 presents the relative biases of the frequencies of the Mediterranean patterns for the ECMWF reforecasts (median value of the bootstraps). As with Figure 5, values are presented with points (and connected with bold lines), when biases are statistically significant (90% two-tailed confidence interval this time). Even for short lead times, some patterns exhibit (significant) biases over 5%. One explanation is that the patterns on ERA5 and ECMWF forecasts are derived from data with slightly different configurations. The patterns on ERA5 are based on daily data derived by averaging hourly data. On the other hand, for ECMWF reforecasts, as the data are not in hourly resolution for all lead times, daily data were based on the average of only 2 values (start and end of day), meaning that a lot of the 24-hours variability and convective activities during daytime are lost.

The main significant biases observed are an underestimation of the Sicilian Low and an overestimation of the Minor Low. These biases are stable for all lead times and mainly occur in summer half. The significant negative biases for the first few lead days of summer half for the Mediterranean pattern, are mainly related to the very low frequency of this pattern during this period (Fig. 2). This leads to high differences in percentages, with only a minor difference in the actual occurrences. During winter half, there is also a consistent overestimation of the Balkan Low and Black Sea Low, mainly significant for most lead times. Nevertheless, most biases (significant and non-significant) lie within the ±10% band, especially for winter. Good representation of the patterns at this period is crucial for indirectly informing about EPEs, as most of the EPEs occur during winter half.

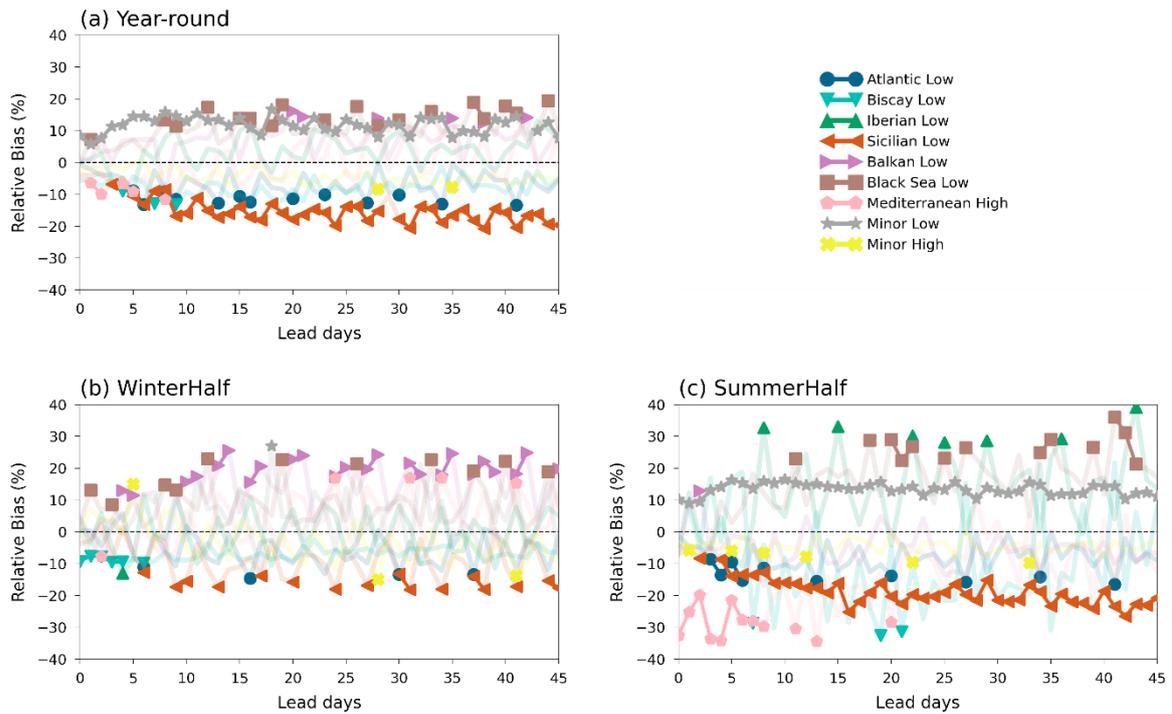

*Figure 6 Relative frequency biases of the Mediterranean patterns for the ECMWF reforecasts (median bootstraps value), when considering annual (a), winter-half (b), and summer-half (c) year periods. The points (connected with bold lines) indicate significant biases.*

The forecast performance (median value of the bootstraps) in the predictability of P95 EPEs is presented in Figure 7 (Figs. S6, S7 for P90, P099, respectively). The plot shows the performance over selected (large) domains (area-weighted mean of all included grid cells), in terms of BSS for P95 EPEs, based on direct forecasting (using model-predicted precipitation) and indirect one (using the forecasted patterns and their half-year conditional probabilities for EPEs). The direct forecasts have a high skill for short and medium range predictions; the skill drops below the reference score (pale-coloured lines) at about 8 days lead time. A small spike observed at 15 days lead time can be attributed to the change of the horizontal resolution that happens to the ECMWF forecasts at that time, which lead to some errors when resampling the variable from the fine resolution to the coarser one. The indirect EPE forecasting outperforms the reference score over 10 days lead time in general, extending the forecasting horizon. This extension of the forecasting horizon based on indirect forecasting of EPEs can be explained by the nature of this method, which, by construction, is not biased. More specifically, when there are no (or low) biases on the regimes (something that in general applies in this study; Fig. 6), the indirect method does not over-/under-forecast EPEs, and that's why the BSS has an asymptotic behaviour around 0 at longer lead times. On the other hand, the direct forecasting of precipitation is more susceptible to biases, especially so at longer lead times. Thus, BSS decreases to below 0 after medium-range forecasts. As the discriminatory skill of the patterns nowhere exceeds 20% even for perfect forecasts (Fig. 4), the BSS does not exceed 0.2, even at 0-day lead time. From the selected subregions (subplots b-f), northern Morocco has a better skill. This relates to its small domain and homogeneity, with the mountain ranges forming a barrier that direct the flow and force moisture to precipitate.

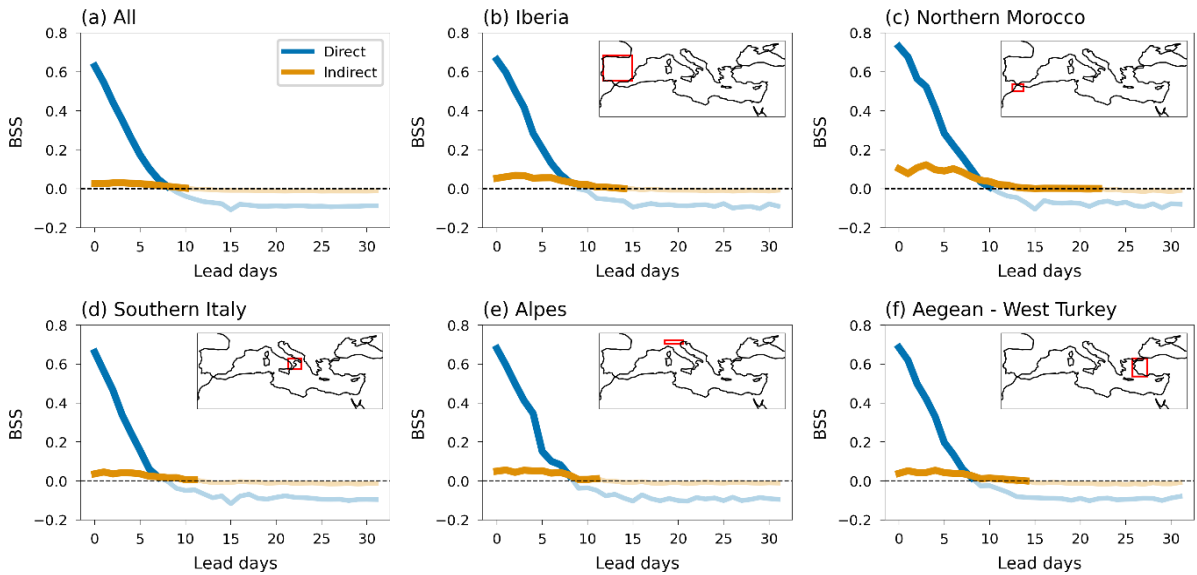

*Figure 2 Brier Skill Score for direct (blue colour) and indirect (orange colour) P95 EPEs forecasting for selected subregions.*

Figure 8 presents similar results as Figure 7, yet now for each grid cell of the studied domain. Subplots (a) and (b) present the forecasting horizon up to when the model beats the reference score for direct and indirect forecasting (statistically significant results). The results of the indirect forecasting are masked to exclude any grid cell that has no significant discriminatory skill for EPEs when conditioning on the patterns (i.e., assuming a perfect pattern forecast). Significance is derived based on the 1,000 bootstraps per lead day (and considering all lead times used for the analysis) with a 90% one-tailed confidence interval. The indirect forecasting outperforms the reference score even for over 8 days lead time for many regions, especially so for Iberian Peninsula, southern Balkans and western Turkey. In contrast, direct forecasting has skilful predictions only up to 8 days lead time for most locations. It can be noticed from subplot (c) that indirect forecasting extends the forecasting horizon for EPEs by between 3 and 6 days for many of the locations. This is a substantial increase in lead time, which could support informative decision making for various domains, as for example (re)insurance companies and agricultural sector. As the discriminatory skill (Fig. 4) of the patterns for inferring EPEs is not very strong, indirect forecasting is not beneficial at short lead times. This information outperforms the direct forecasting only after the end of week 1 for most locations as shown in subplot (d).

Summarizing, results suggest that up to the end of week one, it is beneficial to use direct EPE forecasting, while for week two (extending to week three for few locations), indirect forecasting is the preferred method. Therafter, climatological forecasting provides the most informative inputs. Results for P90 and P99 events are presented in Figures S8 and S9. As the EPEs definition becomes stricter, benefits of indirect forecasting decrease and are confined to smaller areas; mainly related to locations of high orography and coastal domains. One explanations for the above have been provided in the discussion of Figure 4. An additional explanation is that P99 EPEs refer to a very small sample of events per domain. Thus, conditional probabilities are not stable and deviate noticeable when using bootstrapping and different subsets. This could be improved by optimizing the probabilities. Yet, significant improvement is not expected, as many events are driven by localized convective activity for high EPEs, which is not substantially affecting large-scale weather variability.

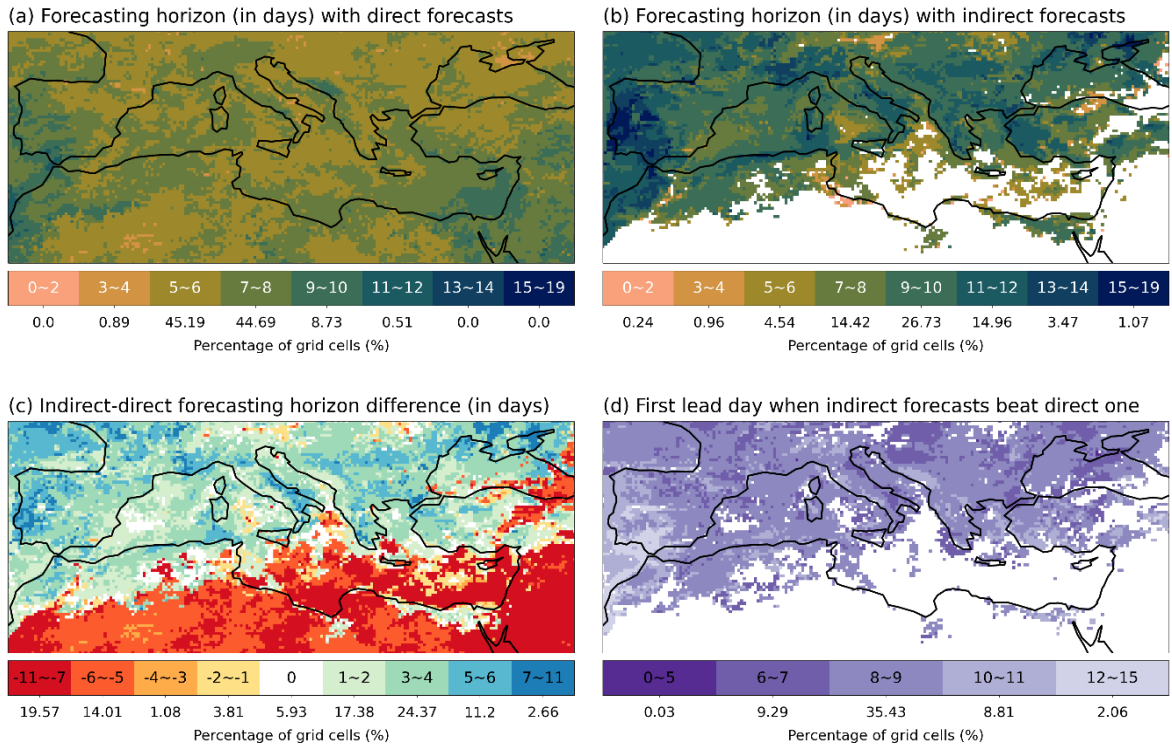

*Figure 8 Forecasting horizon (maximum forecast day) up to when the ECMWF model outperforms the reference score (BSS>0) in predicting P95 EPEs, when assessing EPEs based on the forecasted precipitation (a), or the forecasted Mediterranean patterns (b). (c) Difference of the forecasting horizon between subplot (b) and subplot (a). (d) Minimum lead day from when indirect forecasting of EPEs (based on patterns) outperforms direct forecasting (based on precipitation).*

Figure 9 presents the maximum economic value for 7-, 11-, and 17-days lead-time forecasts, for the same regions as Figure 7. Results are calculated from the actual reforecasts data (using 105 instead of 110 dates, to respect the climatological frequency of seasons), without performing bootstraps due to the computational load. For each of the lead times, the best forecasting method (direct, indirect, reference) for each cost-loss ratio, is selected and coloured accordingly. Figure S10 provides an example of how each of the 3 methods performs for the 7-days lead time. From this plot, the best method at each cost-loss ratio is kept for deriving the 7-day lead-time lines in Figure 9. Note that the reference score has positive values, rather than 0, because it already takes into consideration the seasonality in the occurrence of EPEs, whereas the climatological economic value is solely based on the 5% probability of the EPEs throughout the year. As the studied EPEs are the P95, the maximum economic value is obtained when the cost-loss ratio is equal to the climatological occurrence of the extremes (5 %). For high cost-loss ratios, even for extended-range forecasts, the highest benefits are obtained when using model-predicted precipitation. This is because for such high ratios, decisions would be financially viable only if there is high probability and confidence for upcoming EPEs. Direct forecasting is the only method that can deliver high probabilities, as it can even reach 100 %, whereas indirect (or the reference ones), cannot exceed 45 % at any location (Fig. 2). As for low cost-loss ratios, indirect forecasting brings the highest benefits, already from day 7. The usefulness of this information becomes wider at 11 days lead time, while benefits can be seen even 17 days ahead. These results further support the previous findings in terms of BSS, translating the usefulness of the forecasts to economic benefits. They demonstrate that there is a wide range of cost-loss ratios that indirect forecasting based on the patterns delivers maximum benefits for decisions regarding EPEs.

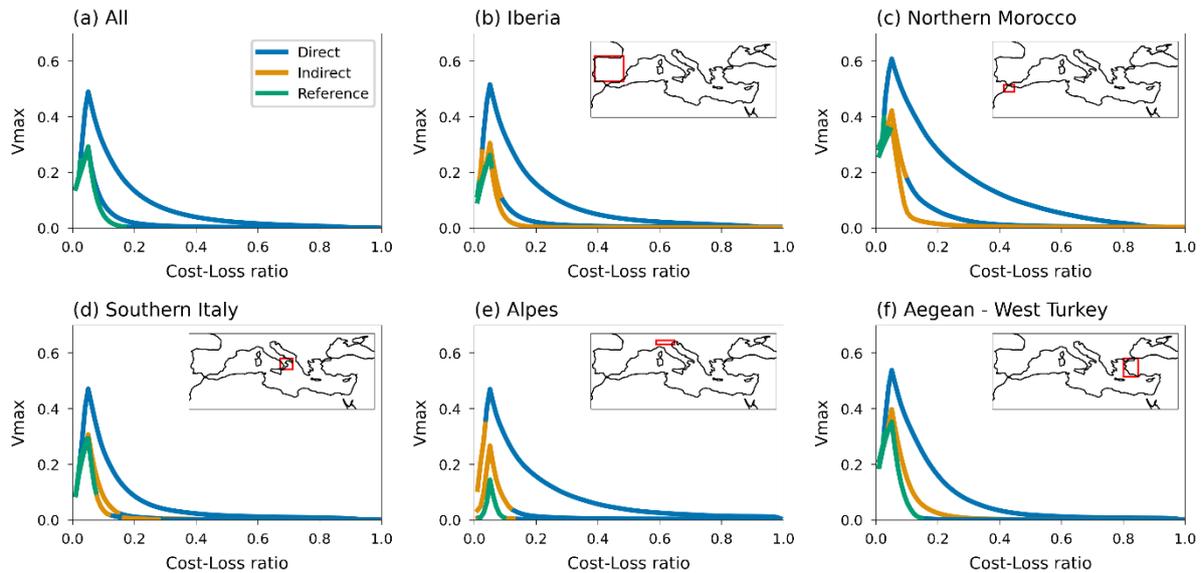

*Figure 9 P95 EPEs: Maximum Economic value for all cost-loss rations and three different lead times (7-, 11-, and 17-days from upper line to lower line) for selected subregions. At each cost-loss ratio, the best method among the direct/indirect forecasting and reference score are kept and coloured accordingly (blue, orange, green colour respectively).*

## 5. Conclusions

This work analysed if and where large-scale atmospheric flow patterns over the Mediterranean can be used as skilful predictors for localized EPEs at medium and extended range forecasts. The nine selected large-scale patterns by Mastrantonas et al. (2021), based on EOF analysis and subsequent K-means clustering of daily anomalies of sea level pressure and geopotential height at 500 hPa, depict atmospheric variability in the lower and middle troposphere over the Mediterranean. The EPEs were derived from the 90$^{th}$, 95$^{th}$, and 99$^{th}$ percentile of annual daily precipitation at each grid cell. The ERA5 dataset was used as the reference dataset, while the ECMWF extended range reforecasts (cycle 46r1) were used as forecasting product. Long-term statistics of the product, regarding pattern predictability and indirect EPE predictability, based on the predicted patterns, was assessed with the Brier Skill Score, considering 2,200 reforecasts at each lead time ranging from 0 up to 45 days ahead. Bootstrapping was also implemented to assess results' significance.

The results show that the ECMWF model well represents the Mediterranean patterns without remarkable biases of their climatological frequencies – even up to 45$^{th}$ day lead time. The model provides skilful predictions of the patterns up to 2 weeks in advance; outperforming results based on climatological frequencies and persistence. Its performance does not show noticeable deviations between the different patterns and the winter-half and summer-half periods. The only differences worth-noticing are observed during winter-half years for Mediterranean High and Minor Low patterns. The former has a forecasting horizon extending up to week 3, while the latter is limited to week 1.

Using the forecasted patterns for indirect EPE predictability provides skilful predictions up to ~10 days for many locations in the Mediterranean. Especially for areas with high orography and coastal locations (e.g., parts of Iberian Peninsula, Morocco, western Italy/Balkans/Turkey), the use of these patterns outperforms climatological EPE allocations by more than 10 days ahead. In fact, using these patterns, rather than the actual forecasted precipitation fields, extends the EPE forecasting horizon by 3–6 days for many locations. These benefits also translate to economic value, with results indicating that the maximum economic benefits for low cost-loss ratios are obtained based on indirect EPE forecasting from 7-days lead time already.

Our results demonstrate that using large-scale patterns as predictors can provide useful information for localised extremes already at medium-range forecasts, extending up to sub-seasonal scales. Such information can be promising for various users, for example the agricultural sectors, emergency response units and (re)insurance companies. To further advance this direction, it would be useful to research additional aspects:

- Are there teleconnections influencing the occurrence of the nine Mediterranean patterns?
- Are there sources of predictability that give higher confidence for the forecasts of the patterns at extended-range lead times?
- What would be the benefits of using more localised large-scale patterns for different subdomains in the Mediterranean, considering, for example, country-wise analysis?
- How skilful is indirect EPE forecasting when using other predictors, such as water vapour flux that is highly related to precipitation (Lavers et al., 2018, 2017, 2016a, 2016b, 2014, 2011)?

Answering such questions can provide guidelines about which predictor is beneficial for different spatiotemporal resolutions, locations, and forecasting horizons, making better use of already available NWP model outputs. This can ultimately support the development of new operational products towards seamless predictions of extreme precipitation that will provide higher confidence to decision makers and users of different sectors. Such steps, considered a priority for international research (Majumdar et al., 2021), will be addressed by future studies.

## Acknowledgements

This work is part of the Climate Advanced Forecasting of sub-seasonal Extremes (CAFE) project; a project funded by the European Union's Horizon 2020 research and innovation programme under the Marie Skłodowska-Curie grant agreement No 813844. We would like to thank the data providers for making the data freely available; ERA5 is available through the Copernicus Climate Data Store and ECMWF reforecasts were downloaded via the ECMWF MARS server. NM would like to thank Zied Ben Bouallègue for the discussions regarding forecasts verifications and David Richardson for providing useful comments that improved the quality of this work.

## Author Contributions

N.M. conducted literature review, data analysis and wrote the initial manuscript. L.M., F.P., and J.M. followed the work and reviewed earlier versions. All authors contributed to discussions throughout the development of this study and improved the initial manuscript. All authors have read and approved the content of the manuscript.

# Supplementary

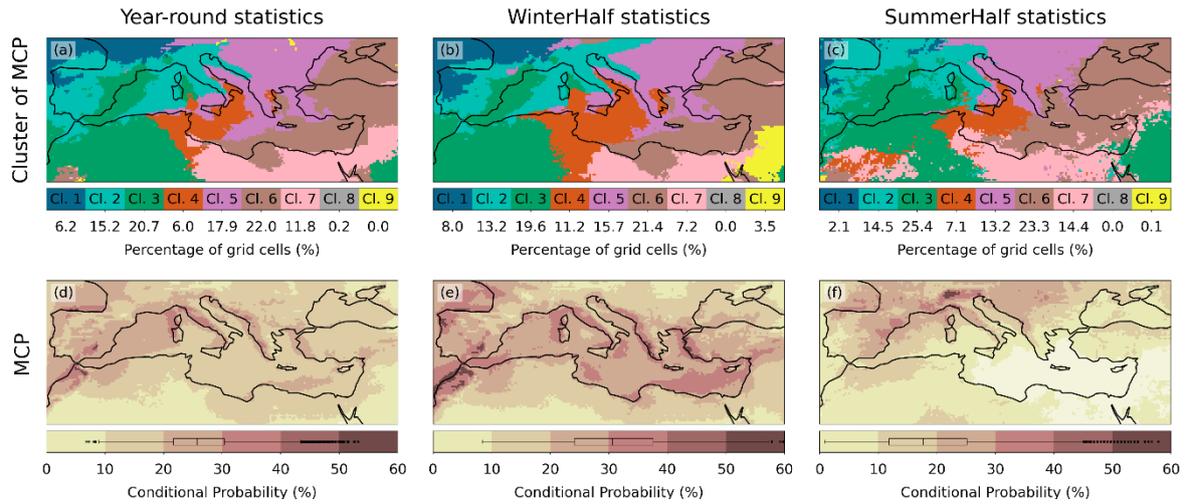

*Figure S1 As Figure 3, but for P90 EPEs.*

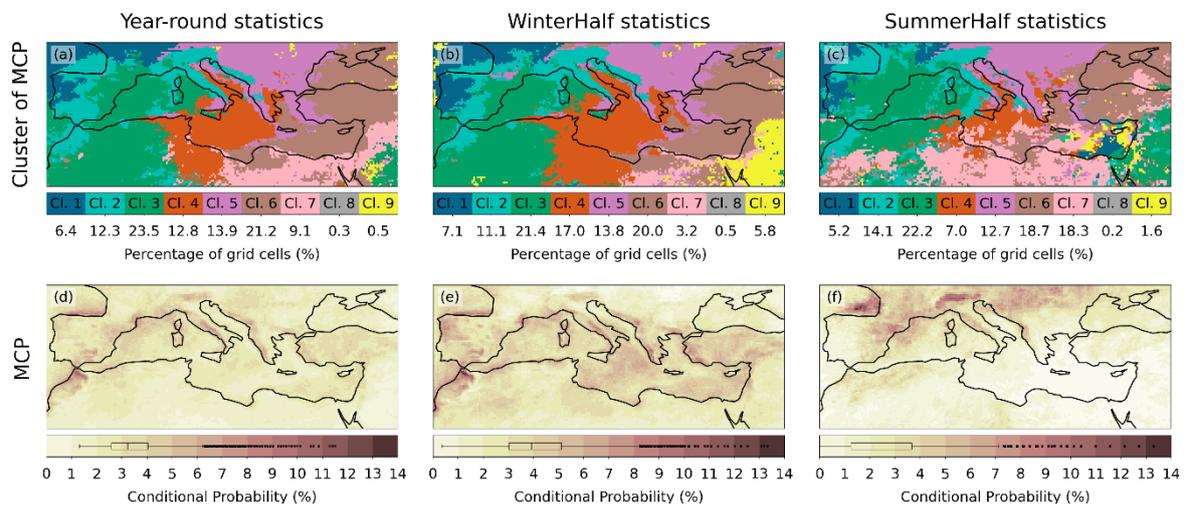

*Figure S2 As Figure 3 but for P99 EPEs.*

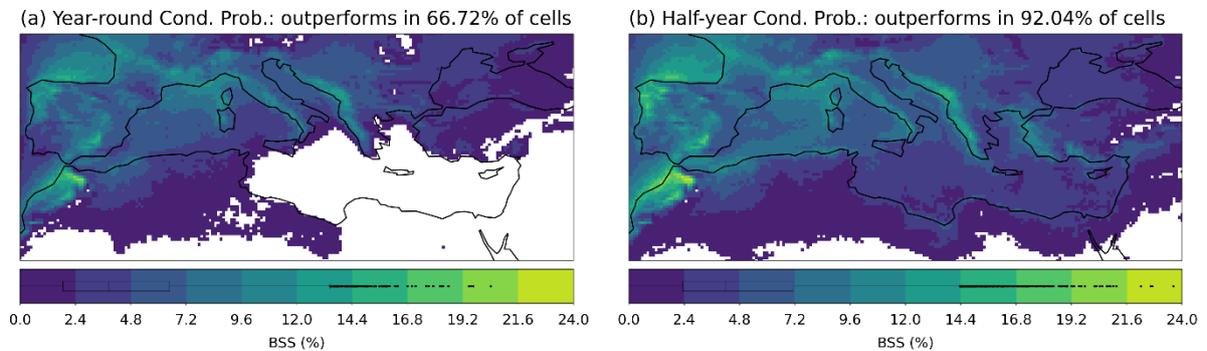

*Figure S3 As Figure 4, but for P90 EPEs.*

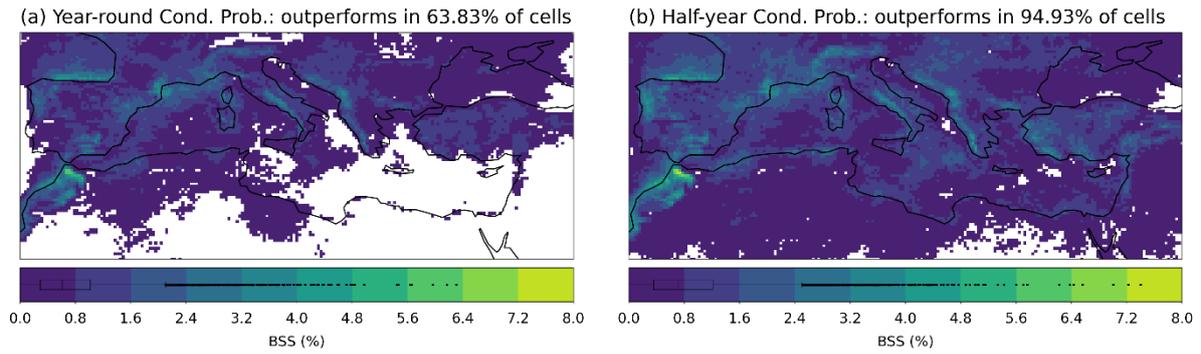

*Figure S4 As Figure 4, but for P99 EPEs.*

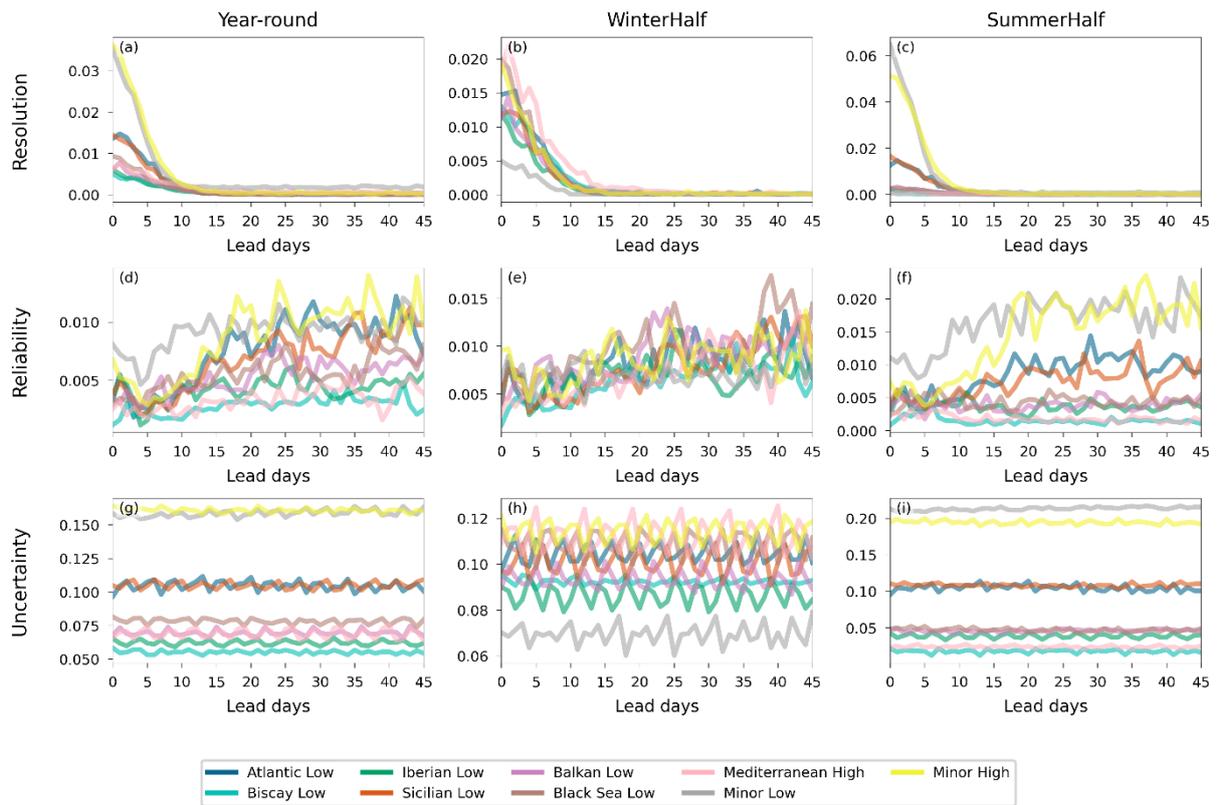

*Figure S5 Brier Score decomposition for the forecasting of the 9 Mediterranean patterns. The 1st column refers to annual statistics, and the 2nd and 3rd to winter- and summer-half years, respectively. The 1st row refers to Resolution, 2nd to Reliability and 3rd to Uncertainty. Note that the y-axes have different scales.*

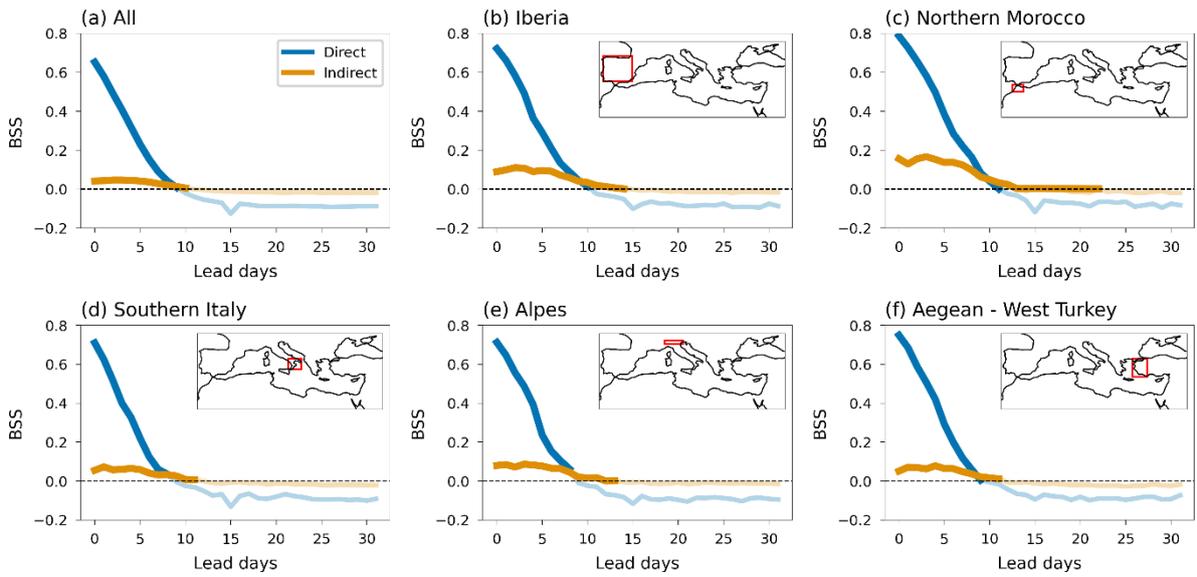

*Figure S6 As Figure 7 but for P90*

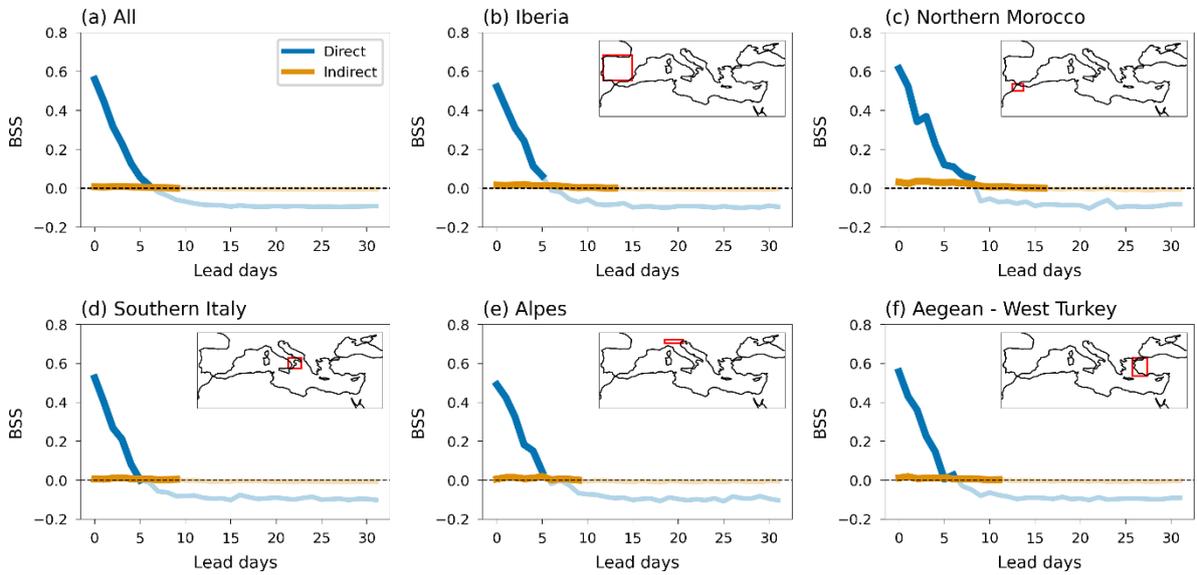

*Figure S7 As Figure 7, but for P99*

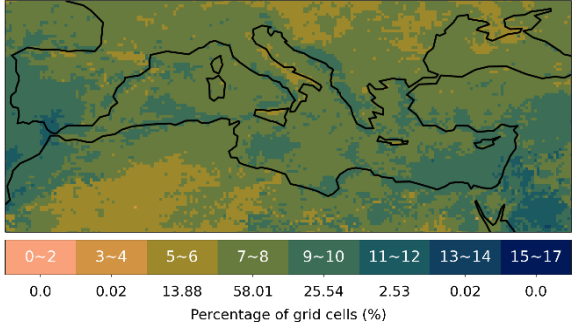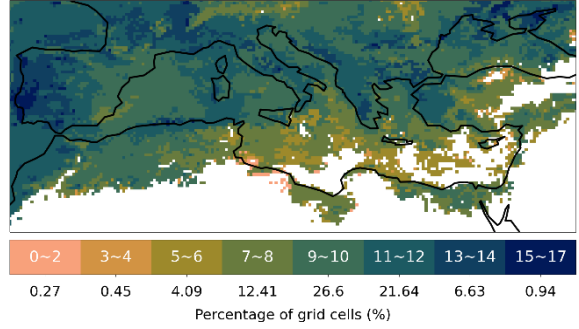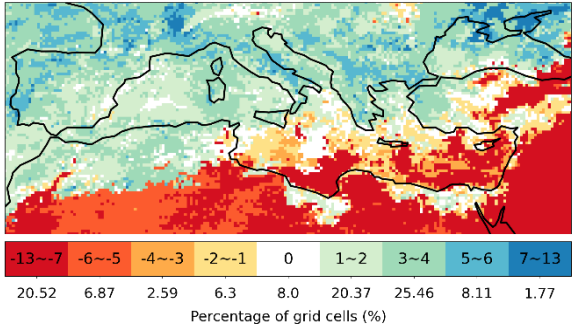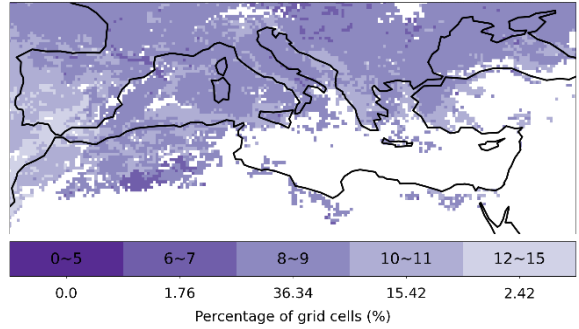

*Figure S8 As Figure 8, but for P90*

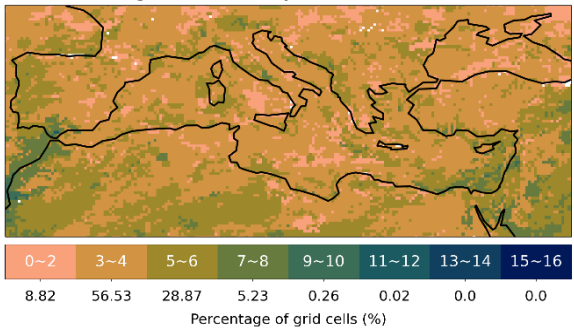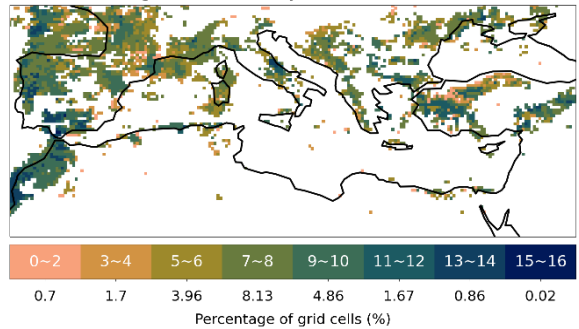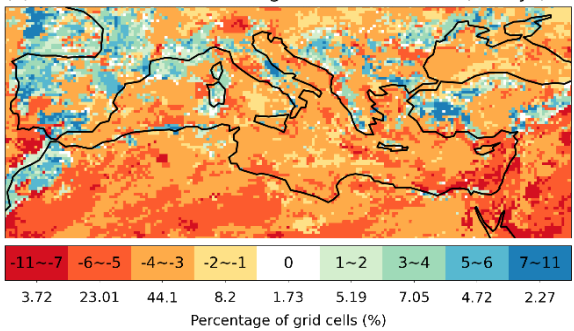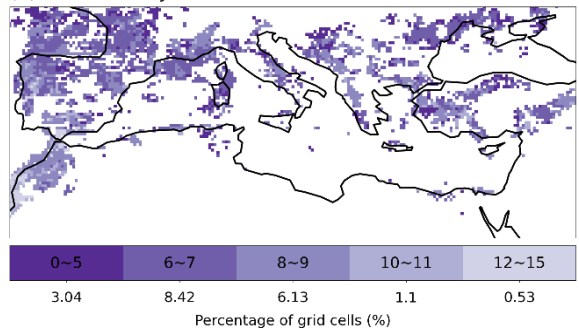

*Figure S9 As Figure 8, but for P99*

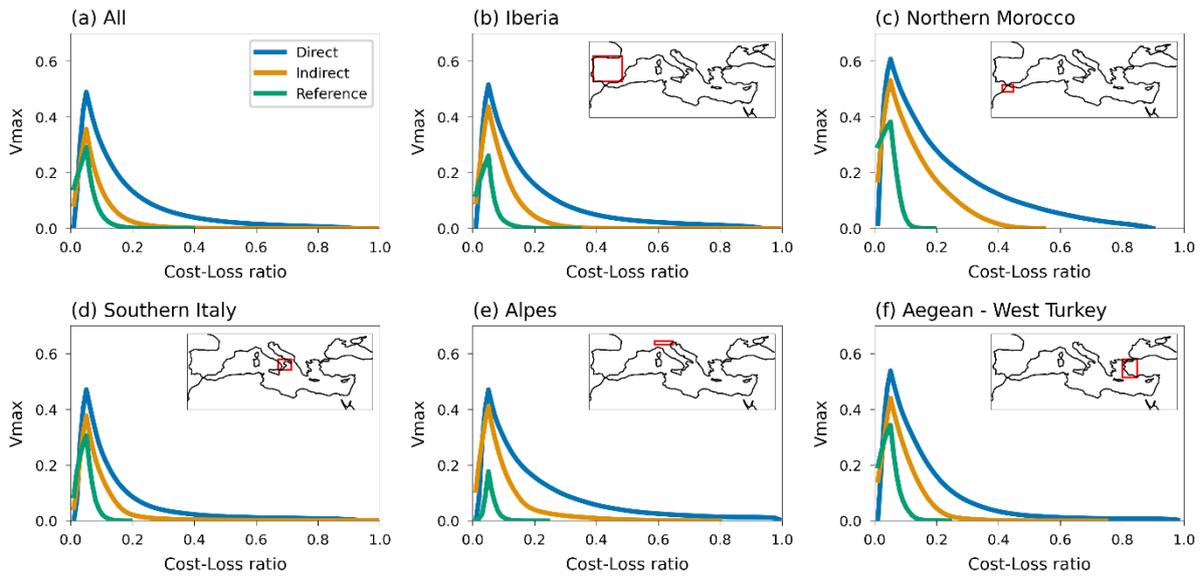

*Figure S10 Maximum Economic Value for P95 EPEs for each forecasting method*

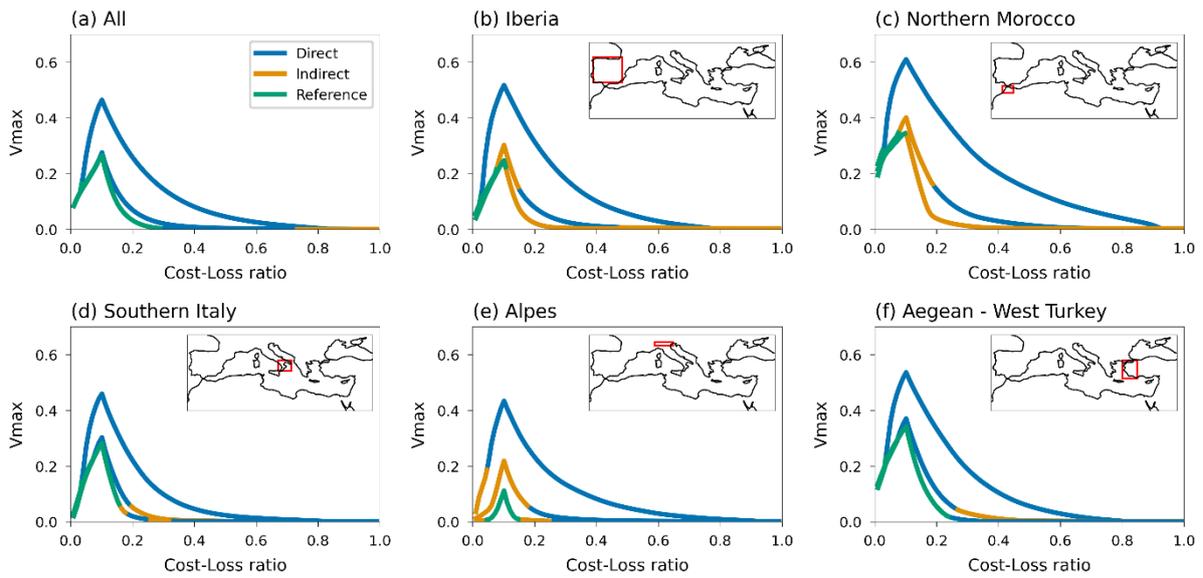

*Figure S11 As Figure 9, but for P90*

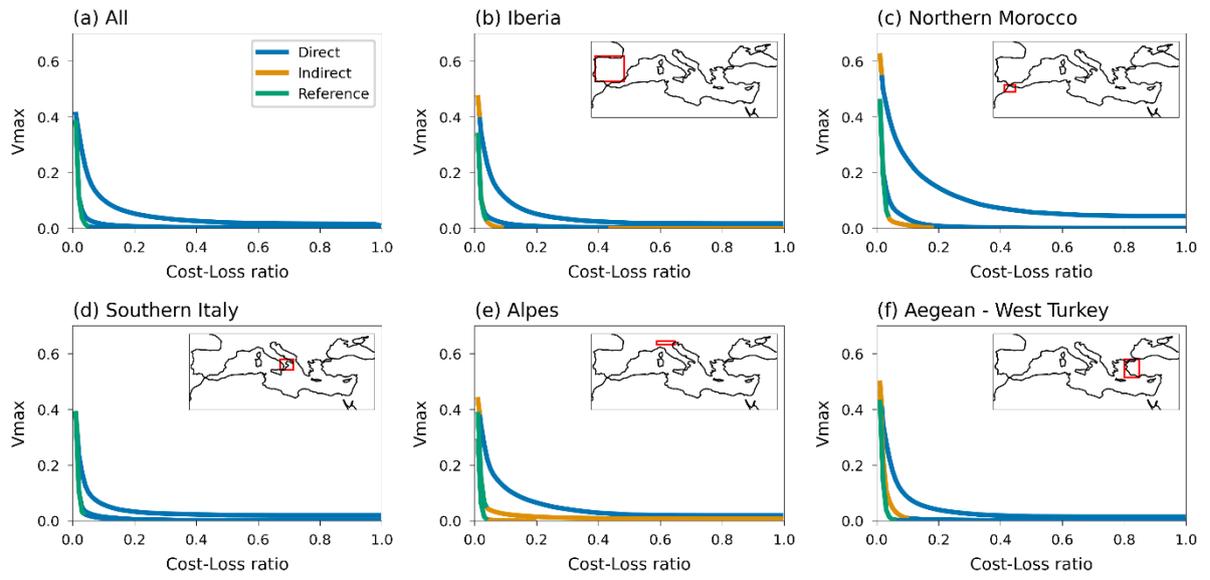

*Figure S12 As for Figure 9, but for P99*